\documentclass[aps,prb,twocolumn,showpacs]{revtex4}
\usepackage{graphicx}

\begin{document}

\title{Superconductivity induced by Pd-doping in SrFe$_{2-x}$Pd$_x$As$_2$}

\author{Xiyu Zhu, Fei Han, Peng Cheng, Bing Shen and Hai-Hu Wen}\email{hhwen@aphy.iphy.ac.cn }

\affiliation{National Laboratory for Superconductivity, Institute of
Physics and Beijing National Laboratory for Condensed Matter
Physics, Chinese Academy of Sciences, P. O. Box 603, Beijing 100190,
China}

\begin{abstract}
By using solid state reaction method, we have synthesized the
Pd-doped superconductor SrFe$_{2-x}$Pd$_x$As$_2$. The systematic
evolution of the lattice constants indicated that the Fe ions were
successfully replaced by the Pd. By increasing the doping content of
Pd, the antiferromagnetic order of the parent phase is suppressed
and superconductivity is induced at a doping level of about x=0.05.
Superconductivity with a maximum transition temperature T$_c$ of
about 8.7 K was achieved at a doping level of x = 0.15. The general
phase diagram of T$_c$ versus x was obtained and found to be similar
to the case of Ni and Co doping to the Fe sites. Our results suggest
that superconductivity can be easily induced in the FeAs family by
adding charges into the system, regardless with the transition
metals of 3d or higher d-orbital electrons.
\end{abstract} 
\maketitle

Superconductivity in the FeAs-based systems has received tremendous
attention since last year with the hope that the superconducting
transition temperature could be raised to a higher
value\cite{Hosono}. The superconducting transition temperature was
promoted above 50 K by replacing lanthanum with other rare-earth
elements in LaFeAsO\cite{Ren,Xuza}, or by substituting the alkaline
elements with rare earth elements in
(Ba,Sr,Ca)FeAsF\cite{ZhuXYEPL,ChengPEPL}. Meanwhile, the hole-doped
superconductors were discovered both in the FeAs-1111 and the
FeAs-122 family\cite{wen,Rotter,CWCh}. The FeAs-122 phase, due to
the much simpler structure, less elements in the compound and easy
growth of large scale single crystals, provides us a great
opportunity to investigate the intrinsic physical
properties\cite{Canfield,LuoHQ}. Meanwhile, people found that a
substitution of Fe ions with Co or Ni can also induce
superconductivity with the maximum T$_c$ above 20
K\cite{Sefat,XuZA,BaNiFeAs}. These early experiments were focused on
the substitution of Fe ions with 3d-transition metals nearby Fe.
Very recently, Ru and Ir substitution at Fe sites in the FeAs-122
phase has also been shown to exhibit
superconductivity\cite{BaFe2-xRuxAs2,SrFeIrAs}. Therefore, it seems
intriguing to know whether it is possible to induce
superconductivity by substituting Fe ions with other transition
metals, such as Pd which locates just below Ni in the periodic table
of elements. In this paper, we report the successful fabrication of
new superconductors in SrFe$_{2-x}$Pd$_{x}$As$_{2}$. X-ray
diffraction pattern (XRD), resistivity, DC magnetic susceptibility,
upper critical field as well as the phase diagram have been
determined in the system of SrFe$_{2-x}$Pd$_{x}$As$_{2}$.

We employed a two-step solid state reaction method to synthesize the
SrFe$_{2-x}$Pd$_{x}$As$_{2}$ samples. Firstly, SrAs, FeAs powders
were obtained by the chemical reaction method with Sr pieces, Fe
powders (purity 99.99\%), and As grains. Then they were mixed
together in the formula SrFe$_{2-x}$Pd$_{x}$As$_{2}$, ground and
pressed into a pellet shape. All the weighing, mixing and pressing
procedures were performed in a glove box with a protective argon
atmosphere (both H$_2$O and O$_2$ are limited below 0.1 ppm). The
pellet was sealed in a silica tube with 0.2 bar of Ar gas and
followed by heat treatment at 900 $^o$C for 50 hours. Then it was
cooled down slowly to room temperature. A repeating of above
procedures improves the purity of the samples.

The x-ray diffraction measurement was performed at room temperature
using an MXP18A-HF-type diffractometer with Cu-K$_{\alpha}$
radiation from 10$^\circ$ to 80$^\circ$ with a step of 0.01$^\circ$.
The analysis of x-ray powder diffraction data was done by using the
software of Powder-X,\cite{DongC} and the lattice constants were
derived by having a general fitting (see below). The DC
magnetization measurements were done with a superconducting quantum
interference device (Quantum Design, SQUID, MPMS7). The
zero-field-cooled magnetization was measured by cooling the sample
at zero field to 2 K, then a magnetic field was applied and the data
were collected during the warming up process. The field-cooled
magnetization data was collected in the warming up process after the
sample was cooled down to 2 K at a finite magnetic field. The
resistivity measurements were done with a physical property
measurement system PPMS-9T (Quantum Design) with the four-probe
technique. The current direction was changed for measuring each
point in order to remove the contacting thermal power.

\begin{figure}
\includegraphics[width=8cm]{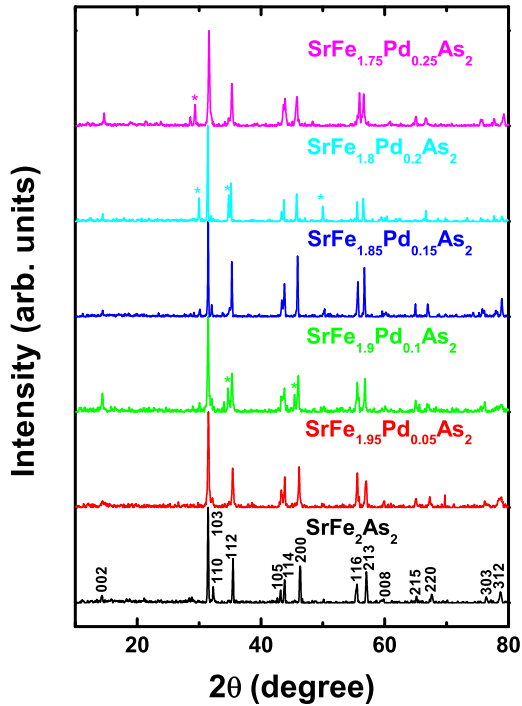}
\caption{(Color online) X-ray diffraction patterns for the samples
SrFe$_{2-x}$Pd$_{x}$As$_{2}$. Almost all main peaks can be indexed
by a tetragonal structure. The asterisks mark the peaks arising from
the impurity phase. } \label{fig1}
\end{figure}

\begin{figure}
\includegraphics[width=8cm]{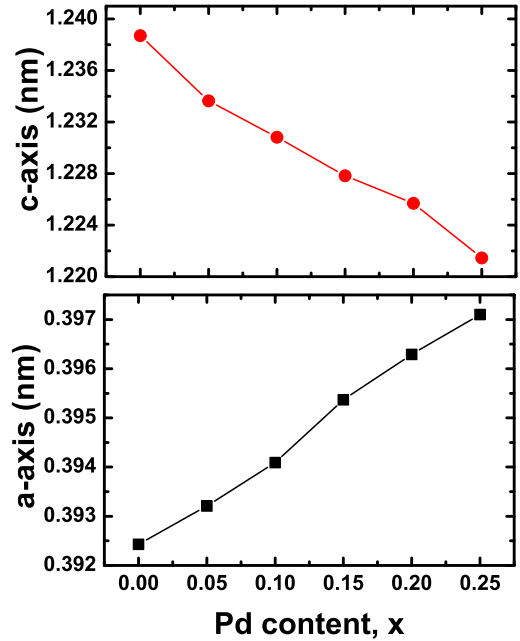}
\caption{(Color online) Doping dependence of a-axis lattice constant
and c-axis lattice constant. It is clear that the a-axis lattice
shrinks, while c-axis lattice expands with Pd substitution. This
systematic evolution clearly indicates that the Pd ions have been
successfully substituted into the Fe-sites. } \label{fig2}
\end{figure}

In order to have a comprehensive understanding to the evolution
induced by the doping effect, we have measured the X-ray diffraction
patterns for all samples. The lattice constants of a-axis and c-axis
are thus obtained. In Figure 1, we present the x-ray diffraction
patterns of SrFe$_{2-x}$Pd$_{x}$As$_{2}$. It is clear that all main
peaks of the samples can be indexed to a tetragonal structure. The
peaks marked with asterisks arise from the impurity phase. By
fitting the data to the structure calculated with the software
Powder-X, we get the lattice constants. In Figure 2, we show a- and
c- lattice parameters for the SrFe$_{2-x}$Pd$_{x}$As$_{2}$ samples.
One can see that, by substituting the Pd into the Fe site, the
lattice constant $a$ shrinks a bit, while $c$ expands slightly. This
tendency is similar to the case of doping the Fe with Ir or Ru in
SrFe$_{2-x}$Ir$_{x}$As$_{2}$ and
BaFe$_{2-x}$Ru$_{x}$As$_{2}$.\cite{BaFe2-xRuxAs2,SrFeIrAs}.

\begin{figure}
\includegraphics[width=8cm]{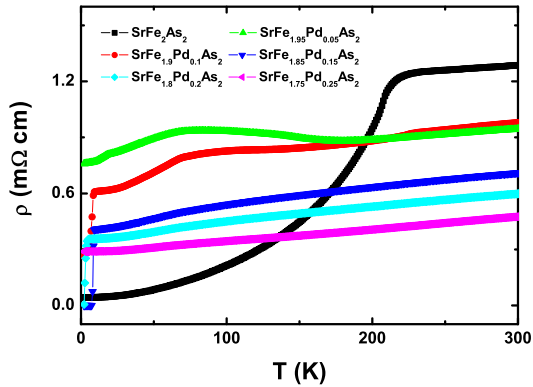}
\caption{(Color online) Temperature dependence of resistivity for
samples SrFe$_{2-x}$Pd$_{x}$As$_{2}$ (x = 0, 0.05, 0.1, 0.15, 0.20
and 0.25, respectively). The superconductivity appears already in
the sample with x=0.10, while the maximum T$_c$ appears at about x =
0.15. } \label{fig3}
\end{figure}

In Figure 3, we present the temperature dependence of resistivity
for samples SrFe$_{2-x}$Pd$_{x}$As$_{2}$ with x = 0, 0.05, 0.1,
0.15, 0.20 and 0.25 respectively. As we can see, the parent phase
exhibits a sharp drop of resistivity (resistivity anomaly) at about
215 K. By doping more Pd, the resistivity drop was converted to an
uprising. This occurs also in the Co-doped samples. We found that
the superconductivity appears in the sample with nominal composition
of x = 0.1. In the sample of x = 0.15, the resistivity anomaly
disappeared completely. This sample shows a superconducting
transition at about 8.7 K which is determined by a standard method,
i.e., using the crossing point of the normal state background and
the extrapolation of the transition part with the most steep slope
(as shown by the dashed lines in Fig.6). The transition width
determined here with the criterion of 10-90 $\%$ $\rho_n$ is about
1.2 K. With higher doping level (x = 0.2) the transition temperature
declines slightly. The superconductivity again disappeared when the
doping content x is over 0.25.

\begin{figure}
\includegraphics[width=8cm]{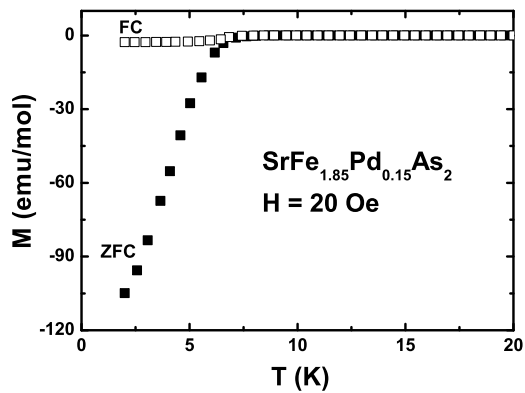}
\caption {(Color online) Temperature dependence of DC magnetization
for the sample SrFe$_{1.85}$Pd$_{0.15}$As$_{2}$. The measurement was
done under a magnetic field of 20 Oe in zero-field-cooled and
field-cooled modes. A strong Meissner shielding signal was
observed.} \label{fig4}
\end{figure}

In Figure 4, the temperature dependence of the DC magnetization for
the sample SrFe$_{1.85}$Pd$_{0.15}$As$_{2}$ was shown. The
measurement was carried out under a magnetic field of 20 Oe in
zero-field-cooled and field-cooled processes. A clear diamagnetic
signal appears below 8.2$\;$K, which corresponds to the middle
transition temperature of the resistivity data. A very strong
Meissner shielding signal was observed in the low temperature
regime, which is similar to the case in
SrFe$_{2-x}$Ir$_{x}$As$_{2}$.

\begin{figure}
\includegraphics[width=8cm]{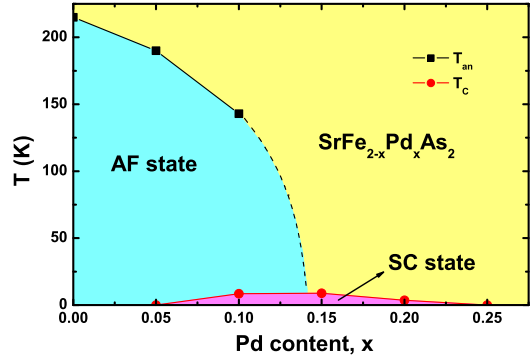}
\caption {(Color online) Phase diagram of
SrFe$_{2-x}$Pd$_{x}$As$_{2}$ within the range of x = 0 to 0.25. The
temperature of resistivity anomaly represents the upturning point of
resistivity which indicates a deviating point from a rough T-linear
behavior in the high temperature region. The superconductivity
starts to appear at x > 0.05, reaching a maximum T$_c$ of 8.7 K at x
= 0.15. The dashed line provides a guide to the eyes for the
possible AF order/strctural transitions near the optimal doping
level.} \label{fig5}
\end{figure}

In Figure 5, a phase diagram of SrFe$_{2-x}$Pd$_{x}$As$_{2}$ within
the range of x from 0 to 0.25 was given. Both T$_{an}$ and T$_{c}$
was defined as temperature of the anomaly in resistivity and the
superconductivity transition by resistivity and susceptibility,
respectively. Like other samples in FeAs-122 phase, with increasing
Pd-doping, the temperature of the resistivity anomaly which may
correspond the tetragonal-orthorhombic structural /
antiferromagnetic transition is driven down, and the
superconductivity emerges at x = 0.1, reaching a maximum T$_c$ of
8.7 K at x=0.15. The superconducting state disappeared at x = 0.25.
As we can see, there exists a region in which the antiferromagnetic
and superconductivity coexist. This general phase diagram looks very
similar to that with Ni-doping.\cite{XuZA} Since Pd locates just
below Ni in the periodic table of elements, we would conclude that
the superconductivity induced by Pd doping shares the similarity as
that of Ni doping. Further measurements on other properties will
clarify this point.

\begin{figure}
\includegraphics[width=8cm]{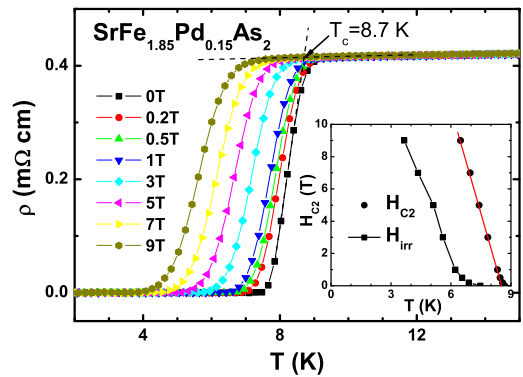}
\caption {(Color online) Temperature dependence of resistivity for
the sample SrFe$_{1.85}$Pd$_{0.15}$As$_{2}$ at different magnetic
fields. The dashed line indicates the extrapolated resistivity in
the normal state. One can see that the superconductivity seems to be
robust against the magnetic field and shifts slowly to the lower
temperatures. The inset gives the upper critical field determined
using the criterion of 90\%$\rho_n$. A slope of -dH$_{c2}$/dT = 4.2
T/K near T$_c$ is found here. The irreversibility line H$_{irr}$
taking with the criterion of 0.1\% $\rho_n$ is also presented in the
inset.} \label{fig4}
\end{figure}

In Figure 6 we present the temperature dependence of resistivity
broadening induced by using different magnetic fields. Just as many
other iron pnictide superconductors, the superconductivity is also
very robust against the magnetic field in the present sample
although the T$_c$ is only 8.7 K. We used the criterion of
$90\%\rho_n$ to determine the upper critical field and show the data
in the inset of Figure 6. Surprisingly, regarding the relatively low
T$_c$, a high slope of -dH$_{c2}$/dT = 4.2 T/K can be obtained here.
By using the Werthamer-Helfand-Hohenberg (WHH) formula\cite{WHH}
$H_{\mathrm{c}2}(0)=-0.69(dH_{c2}/dT)|_{T_c}T_c$, the value of zero
temperature upper critical field can be estimated. Taking
$T_\mathrm{c}= 8.7\;$K, we get $H_{\mathrm{c}2}(0) \approx 25.1 T$
roughly. Because of the low superconducting transition temperature,
the present Pd-doped sample has a smaller upper critical field,
compared with K-doped\cite{WangZSPRB} and Co-doped samples\cite{Jo}.

The superconductivity mechanism in the FeAs-based superconductors
remains unclear yet. One widely perceived picture is that the
pairing is established via the inter-pocket scattering of electrons
through exchanging the AF spin
fluctuations.\cite{Mazin,Kuroki,WangF,WangZD} By doping electrons or
holes into the parent phase, the condition for forming the AF order
will be destroyed gradually. Instead the short range AF order will
provide a wide spectrum of spin fluctuations which may play as the
media for the pairing between electrons. This picture can certainly
give a qualitative explanation to the occurrence of
superconductivity here. However, it is still unclear why the
superconducting transition temperature varies in doping different
elements. For example, the maximum T$_c$ by doping Co, Ni, Ru or Ir
can be as high as 24-26
K,\cite{Sefat,SrFeIrAs,BaFe2-xRuxAs2,BaNiFeAs}, while that by doping
Pd in the present case is only about 8.7 K. In addition, in most
cases, the substitution to the Fe sites by other transition metal
elements in the 1111 phase gives only a rather low superconducting
transition temperature compared to the 122 phase. This puzzling
point certainly warrants further investigations. Our data here
further illustrate that the superconductivity can be easily induced
by doping the Fe sites with many other transition metals which are
not restricted to the ones with 3d orbital electrons.

In summary, superconductivity has been found in
SrFe$_{1-x}$Pd$_{x}$As$_2$ with the maximum T$_c$ = 8.7 K. The phase
diagram obtained is quite similar to that by doping Co, Ni, Ru or Ir
to the Fe sites. The superconductivity is rather robust against the
magnetic field with a slope of -dH$_{c2}$/dT = 4.2 T/K near T$_c$.
Our results clearly indicate that the superconductivity can be
easily induced in (Ba,Sr)Fe$_2$As$_2$ by replacing the Fe sites with
many different transition metal elements which can be those with 4d-
or 5d-orbital electrons.

This work is supported by the Natural Science Foundation of China,
the Ministry of Science and Technology of China (973 project:
2006CB601000, 2006CB921802), the Knowledge Innovation Project of
Chinese Academy of Sciences (ITSNEM).

\end{document}